\documentclass[a4paper,11pt]{article}
\usepackage{pos}
\usepackage{cleveref}
\usepackage{mathtools}
\usepackage[normalem]{ulem}
\usepackage{color}

\Crefname{figure}{Fig.}{Figs.}
\Crefname{equation}{Eq.}{Eqs.}
\creflabelformat{equation}{#2\textup{#1}#3}
\Crefname{section}{section}{sections}
\Crefname{table}{Tab.}{Tabs.}

\def\NT{N_{\tau}}
\def\tf{{\tau_{\text{F}}}}
\def\d{\mathrm{d}}
\newcommand{\stackss}[2]{{\begin{subarray}{l}{\textrm{\tiny #1}}\\[-0.3ex] \textrm{\tiny #2}\end{subarray}}} 

\title{Continuum extrapolation of the gradient-flowed color-magnetic correlator at $1.5\,T_c$}

\author*[a]{Luis Altenkort}
\author[b]{Alexander M. Eller}
\author[a]{Olaf Kaczmarek}
\author[c]{Lukas Mazur}
\author[b]{Guy D. Moore}
\author[d]{Hai-Tao Shu}

\affiliation[a]{Fakultät für Physik, Universität Bielefeld\\
  Universitätsstraße 25, D-33615 Bielefeld, Germany}

\affiliation[b]{Institut für Kernphysik, Technische Universität Darmstadt,\\
Schloßgartenstraße 9, D-64289 Darmstadt, Germany}

\affiliation[c]{Paderborn Center for Parallel Computing, Paderborn University,\\
Pohlweg 51, D-33098 Paderborn, Germany}

\affiliation[d]{Institut für Theoretische Physik, Universität  Regensburg,\\
Universitätsstraße 31, D-93053 Regensburg, Germany}

\emailAdd{altenkort@physik.uni-bielefeld.de}

\abstract{In a recently published work we employ gradient flow on the lattice to extract the leading contribution of the heavy quark momentum diffusion coefficient in the heavy quark limit from  calculations of a well-known two-point function of color-electric field operators. In this article we want to report the progress of calculating the recently derived color-magnetic correlator that encodes a finite mass correction to this transport coefficient. The calculations we present here are based on the same ensemble of quenched gauge configurations at $1.5\, T_c$ that we previously used for the color-electric correlator.}

\FullConference{%
 The 38th International Symposium on Lattice Field Theory, LATTICE2021
  26th-30th July, 2021
  Zoom/Gather@Massachusetts Institute of Technology
}

\begin{document}
\maketitle

\section{Introduction}
The foundation of numerous lattice studies of heavy quark diffusion employs heavy quark effective theory (HQET) \cite{EICHTEN1990511,manohar_wise_2000,grozin2004}, which is based on an expansion of the QCD action in the inverse heavy quark mass $1/M$. 
By considering only the leading term of the expansion in a Kubo-type formula for the momentum diffusion coefficient $\kappa$, one can define a gluonic Euclidean two-point function of color-electric field operators which encodes the leading contribution to this transport coefficient \cite{CaronHuot:2009uh}.
More specifically, it is defined as the $\omega\rightarrow 0$ limit of the corresponding spectral function that, in practice, has to be reconstructed from the discrete Euclidean correlator.
The spectral reconstruction is not only an underdetermined integral inversion problem, but in order to obtain useful estimates for the spectral function one needs high-precision correlator data, which makes it utterly necessary to implement some kind of noise reduction technique on the lattice. 
In recent publications \cite{Altenkort2021,Altenkort:2020axj} we have shown that gradient flow \cite{Luscher:2009eq, Luscher:2010iy}  is a solution which is not restricted to local actions and therefore applicable for ensembles with dynamical fermions.

Around the same time, Bouttefeux and Laine used the expanded action up to $\mathcal{O}(1/M^3)$ to derive the corresponding two-point functions up to $\mathcal{O}(1/M^2)$, which ultimately yield one genuine finite mass correction to the leading-order momentum diffusion coefficient that is suppressed by $\mathcal{O}(T/M)$ \cite{Bouttefeux:2020ycy}. This contribution is encoded in a color-magnetic correlator. In this proceedings article we want to present a high-precision continuum extrapolation of this correlator from large and fine lattices at finite flow times in a hot gluonic medium at $1.5 \, T_c$. A flow-time-to-zero extrapolation of the continuum correlator is not as straightforward as in the case of the color-electric correlator because of its nonzero anomalous dimension \cite{Laine:2021uzs}. We briefly discuss this problem at the end of \Cref{sec:context}, but declare this task to be out of scope of this article. 

\section{Diffusion physics from gluonic correlation functions}

Starting from the standard Kubo-formula for the diffusion coefficient D (see \cite{CaronHuot:2009uh}), one can find an expression for the momentum diffusion coefficient $\kappa$ by assuming that the diffusive motion of heavy quarks in a hot plasma can be described by non-relativistic Langevin equations. In that case one finds a Lorentzian transport peak in the infrared \cite{Petreczky:2005nh}, whose tail encodes the momentum diffusion coefficient in the ordered limit \cite{CaronHuot:2009uh},
\begin{align}
\kappa \equiv \lim_{\omega\rightarrow 0}\lim_{M\rightarrow \infty} \kappa^{(M)} (\omega),
\end{align}
 of the (non-perturbative) force-force correlator,
\begin{align}
    \kappa^{(M)}(\omega) \equiv \frac{1}{3\chi} \int^\infty_{-\infty} \d t \mathrm{e}^{i\omega t} \int_\mathbf{x} \sum_i \left\langle \frac{1}{2} \left\lbrace {\mathcal{F}}^i (t,\mathbf{x}), {\mathcal{F}}^i (0,\mathbf{0})\right\rbrace \right\rangle, \quad {\mathcal{F}}_i = M \d {\mathcal{J}}_i / \d t,
\end{align}
where $\chi$ is the quark number susceptibility; $i$ and $\mathcal{J}_i$ are the spatial directions and heavy quark vector current, respectively. Note that here $M$ is the vacuum mass and does not include a thermal dispersive correction.

Now one has to switch to heavy quark effective theory and use the $1/M$-expansion of the action, truncated at some order, to evaluate the current derivatives through the canonical equations of motion. In the leading order ($1/M$) one obtains a two-point function of color-electric field operators which encodes the leading contribution $\kappa_E$ (again, see \cite{CaronHuot:2009uh} for a derivation, and \cite{Altenkort2021,Brambilla:2020, mayersteudte2021chromoelectric} for recent non-perturbative calculations). For the first correction, which is of order $(1/M^2)$, the contribution is called $\kappa_B$, as it is encoded in a two-point function of color-magnetic field operators \cite{Bouttefeux:2020ycy}. This color-magnetic correlator is defined as
\begin{align}
G_B(\tau) \equiv \frac{\sum_i \mathrm{Re}\mathrm{Tr} \langle U (\beta; \tau) gB_i(\tau) U(\tau;0) g B_i(0) \rangle}{3 \mathrm{Re}\mathrm{Tr} \langle U(\beta;0)\rangle}.
\label{eq:magcorr}
\end{align}
The non-perturbative lattice calculation of this correlator is the main focus of this article.
For the color-magnetic field we employ a naive Euclidean lattice discretization,
\begin{equation}
    \label{ge}
    \begin{aligned}
         B_1=U_2(\mathbf{x})U_3(\mathbf{x}+\hat{2})-U_3(\mathbf{x})U_2(\mathbf{x}+\hat{3}), \\
 B_2=U_3(\mathbf{x})U_1(\mathbf{x}+\hat{3})-U_1(\mathbf{x})U_3(\mathbf{x}+\hat{1}), \\
 B_3=U_1(\mathbf{x})U_2(\mathbf{x}+\hat{1})-U_2(\mathbf{x})U_1(\mathbf{x}+\hat{2}). \\
    \end{aligned}
\end{equation}
$G_B$ encodes (in the infrared limit of its spectral function) the correction $\kappa_B$ which is suppressed by $\mathcal{O}(T/M)$ such that in total one obtains for the heavy quark momentum diffusion coefficient \cite{Bouttefeux:2020ycy}:
\begin{align}
\kappa_\text{tot} \simeq \kappa_E + \frac{2}{3} \langle \mathbf{v}^2\rangle \kappa_B.
\end{align}
Here $\langle \mathbf{v}^2 \rangle $ is the average squared velocity of the heavy quark in a low-energy effective description with the thermally corrected mass $M_\mathrm{kin}$ \cite{Bouttefeux:2020ycy}:
\begin{align}
    \langle \mathbf{v}^2 \rangle \approx \frac{3T}{M_\mathrm{kin}}\left( 1 - \frac{5T}{2M_\mathrm{kin}} \right), \quad M_\mathrm{kin}^2 = M^2 (1 + \mathcal{O}(\alpha_s^{3/2} T/M )).
\end{align}
To obtain $\kappa_B$ one has to find the infrared limit of the corresponding spectral function \cite{Bouttefeux:2020ycy}, 
\begin{equation}
    \label{definition-kappa}
    \lim_{\omega \rightarrow 0} \frac{2 T}{\omega} \rho_B(\omega) = \kappa_B ,
\end{equation}
where $\rho_B(\omega)$ denotes the spectral function of the color-magnetic correlator $G_B$: 
\begin{align}
    G_B(\tau) = \int_0^\infty \frac{{\rm d}\omega}{\pi} \rho_B(\omega)\frac{\cosh \big{(}\omega (\tau-\frac{1}{2T}) \big{)} } {\sinh \big{(} \frac{\omega}{2T} \big{)} }.
    \label{relation}
\end{align}
In continuum perturbation theory the leading term of $G_B$ turns out to be identical to the one of $G_E$ \cite{Bouttefeux:2020ycy}.
Same as in \cite{Altenkort2021}, we want to normalize our numerical non-perturbative lattice calculations of $G_B$ using the (nonflowed) leading-order perturbative \textit{lattice} correlator of $G_B$, which also turns out to be identical to the one computed for $G_E$. As usual it is divided by $g^2 C_F$, and we cite it here for reference (for unexplained notation see \cite{Altenkort2021}):
\begin{align}
&G^\stackss{norm}{latt}_{\tf=0}(\tau)\equiv - \frac{1}{a^{3} \beta} \sum_{n=-\frac{\beta}{2}}^{\frac{\beta}{2}-1} \cos\left( 2\pi n \tau T \right) &\times\left[1 - \hspace{-4pt} \int\limits_{0}^{\infty} \hspace{-0.5ex}\mathrm{d}x ~e^{- x  \sin^{2}\left(\frac{\pi n}{\NT}  \right) } e^{-\frac{3}{2} x}  \left( I_{0}({\scriptstyle \frac{x}{2}})\right)^{2} 
\left\lbrace   I_{0}({\scriptstyle \frac{x}{2}}) - I_{1}({\scriptstyle \frac{x}{2}}) \right\rbrace \phantom{\int\limits_{ }^{} \hspace{-1.0em}} \right]. 
\label{eq:normcorr}
\end{align}

\label{sec-1}

\section{Continuum extrapolation of the color-magnetic correlator}\label{sec:context}

\begin{table}[b]
    \centering
    \begin{tabular}{ccrcccc}                            
    \hline \hline
    $a$ (fm) & $a^{-1}$ (GeV) & $N_{\sigma}$ & $N_{\tau}$ & $\beta$ & $T/T_{c}$ & \#conf.\tabularnewline
    \hline
    0.0215 & 9.187 & 80 &  20  & 7.0350 &  1.47  & 10000 \tabularnewline
    0.0178 & 11.11 & 96 &  24  & 7.1920 &  1.48  & 10000 \tabularnewline
    0.0140 & 14.14 & 120 & 30  & 7.3940 &  1.51  & 10000 \tabularnewline
    0.0117 & 16.88 & 144 & 36  & 7.5440 &  1.50  & 10000 \tabularnewline
    \hline \hline
    \end{tabular}
    \caption{Lattice spacings, lattice dimensions, $\beta$ values, temperature and number of configurations used for the calculations in this work. The ensemble is identical to the one used in \cite{Altenkort2021}. The lattice spacing $a$ is determined via the Sommer scale $r_0$~\cite{Sommer:1993ce} with parameters taken from~\cite{Francis:2015lha} and updated coefficients from~\cite{Burnier:2017bod}. We use $r_0T_c=0.7457(45)$~\cite{Francis:2015lha}. }
    \label{tab:lattice_setup}
\end{table}

\begin{figure}[t]
    \centering
    \includegraphics[width=0.49\textwidth]{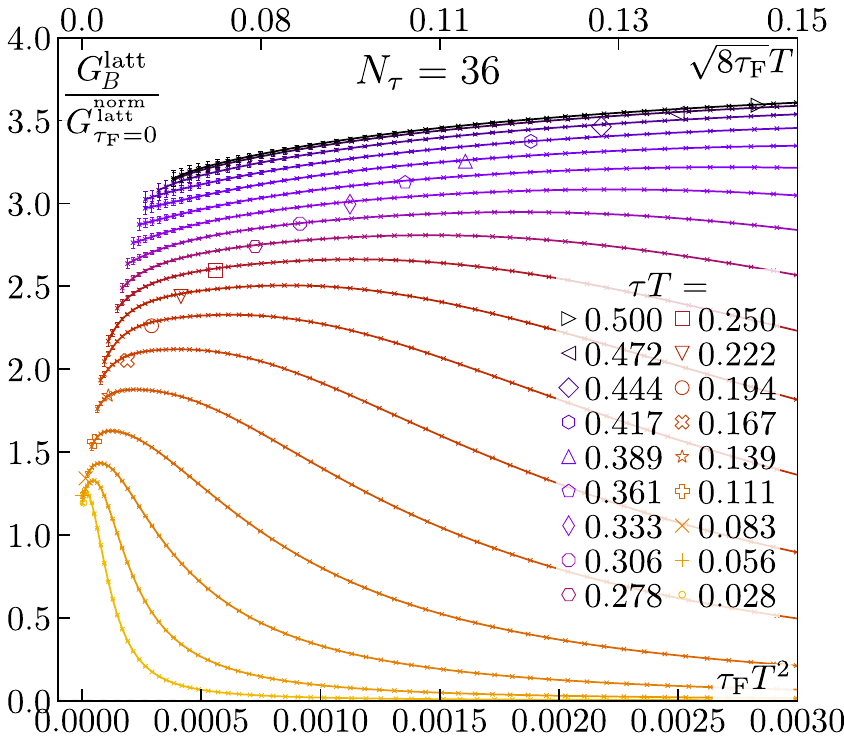}\hfill
    \includegraphics[width=0.49\textwidth]{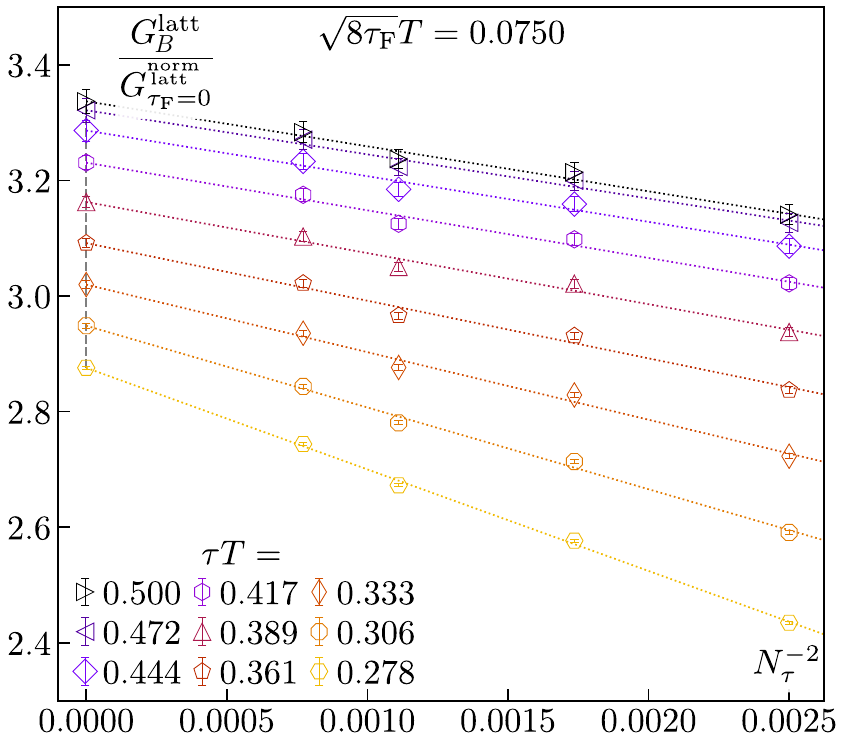}
    \caption{Left: Normalized color-magnetic correlator as a function of flow time at fixed separations $\tau T$. The symbols help to identify the corresponding $\tau T$ to each data curve, and in addition depict a perturbatively calculated upper limit for the flow time, for which the leading-order continuum flowed color-\textit{electric} correlator deviates less than $1\%$ from its nonflowed counterpart \cite{Eller:2018yje}. Data points are connected by straight lines to guide the eye. Right: Continuum extrapolation of the normalized color-magnetic correlator at one intermediate flow time. Here the symbols play no special role other than identifying the $\tau T$. The extrapolated values are connected vertically with a dashed grey line and the bootstrap mean of the linear fit is indicated with dotted colorful lines.}
    \label{fig:BB_Nt36}
\end{figure}

\begin{figure}[t]
    \centering
    \includegraphics[width=0.49\textwidth]{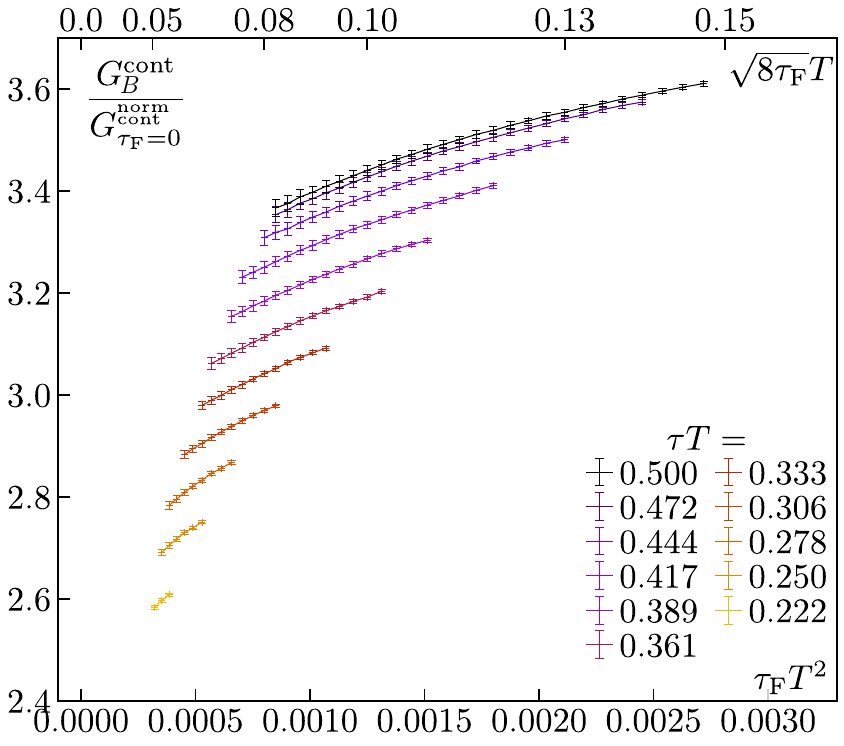}\hfill
    \includegraphics[width=0.49\textwidth]{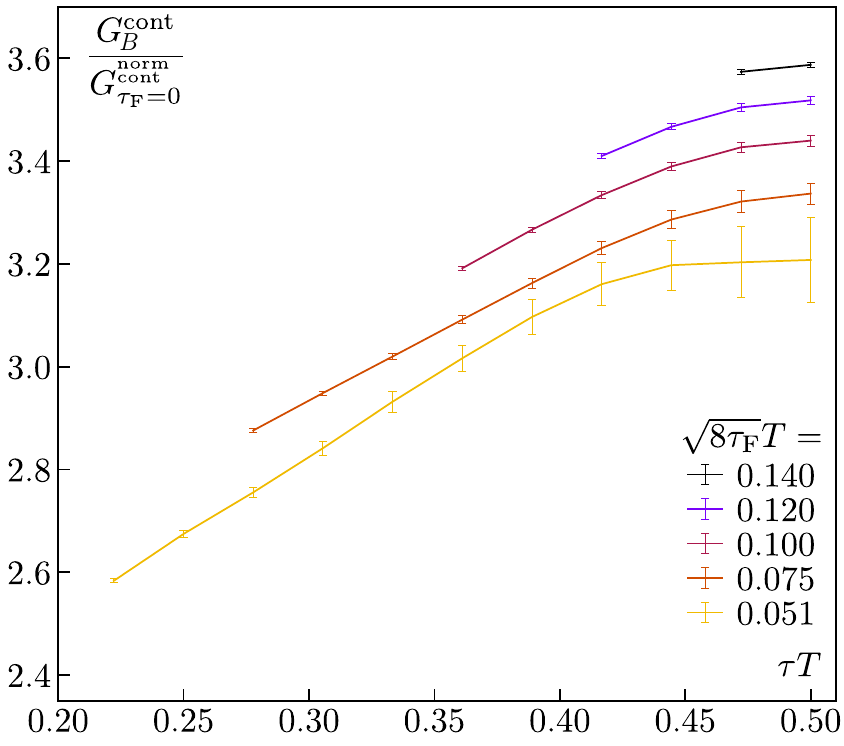}
    \caption{Left: Continuum-extrapolated color-magnetic correlator as a function of flow time at fixed separation $\tau T$. A linear extrapolation in flow time is not applicable. Data points are connected by straight lines to guide the eye. Right: The same as on the left but now as a function of separation at various fixed flow times. Data points are connected by straight lines to guide the eye.}
    \label{fig:BBcontquality}
\end{figure}

In the following we will show the results of nonperturbative calculations of the color-magnetic correlator (\Cref{eq:magcorr}) under gradient flow on the lattice, normalized to its leading-order perturbative counterpart (\Cref{eq:normcorr}).  In the figures we use the notation $G_B^\mathrm{latt}$ for the lattice color-magnetic correlator, and $G_B^\mathrm{cont}$ for the continuum-extrapolated one.

The lattice ensemble is identical to that of \cite{Altenkort2021} and can be found in \Cref{tab:lattice_setup}. The gauge action is the standard Wilson action. For further details, also concerning the numerical integration of the gradient flow equation, we refer to \cite{Altenkort2021}. 
The continuum limit should only be taken on data that is devoid of certain discretization-induced renormalization issues. In other words, we want to keep corrections proportional to $a^2/\tf$ small, which in practice means that the smoothing radius of the flow should be at least one lattice spacing of the coarsest lattice: $\sqrt{8\tf} \approx a$.

The left panel of \Cref{fig:BB_Nt36} shows the color-magnetic correlator as a function of flow time on the finest lattice. The overall magnitude of the correlator is similar but slightly lower compared to the color-electric correlator at the same temperature (cf.\ \cite{Altenkort2021}).
The symbols here depict perturbatively calculated upper limits for the flow time, up to which the leading-order flowed continuum color-\textit{electric} correlator deviates less than $1\%$ from its nonflowed counterpart. We argue that the perturbative calculations for the color-electric correlator can also serve as a first estimate for the color-magnetic one, as their leading-order terms are identical at vanishing flow time \cite{Bouttefeux:2020ycy}. Up to these limits the flowed correlator data should still contain the correct (infrared) physics.
Similar to the color-electric correlator we observe an initial rising behavior that is followed by a region of small to moderate flow time dependence. In the very end all meaningful physics are destroyed and the correlator tends to zero. 

On the right panel of \Cref{fig:BB_Nt36} we show the continuum extrapolation at one moderate flow time ($\sqrt{8\tf}T=0.075$), where the points of the coarser lattices have been obtained by interpolation with cubic splines as in \cite{Altenkort2021}. The leading discretization errors of the Wilson action are of order $a^2$ which is why the continuum extrapolation is a weighted linear fit of the data to 
\begin{align}
    \frac{G^\textrm{latt}_{\tau,\tf}(N_\tau) }{G_{\tau,\tf=0}^{\begin{subarray}{l}\mathrlap{\textrm{\tiny  norm}}\\[-0.4ex] \textrm{\tiny latt}\end{subarray}}(N_\tau)}
     =  m \cdot N_\tau^{-2} + b ,
\end{align}
where $m$ and $b \equiv G_{\tau,\tf}^\mathrm{cont}/G_{\tau,\tf=0}^{\begin{subarray}{l}\mathrlap{\textrm{\tiny  norm}}\\[-0.4ex] \textrm{\tiny cont}\end{subarray}}$ are the parameters to fit. 

In \Cref{fig:BBcontquality} we finally show the continuum-extrapolated color-magnetic correlator: on the left as a function of flow time, and on the right as a function of separation at some fixed flow times. 
In the flow time region that could in principle be used for a flow-time-to-zero extrapolation we do not observe a linear dependence as in the case of the color-electric correlator. The reason for this is that, in contrast to the color-magnetic correlator, the color-\textit{electric} correlator has a vanishing anomalous dimension and so in principle it is independent of the renormalization scale ($\sim \tau_F^{-1}$) at which it is evaluated. In practice there is of course still a flow time dependence: in the window that is suitable for a flow-time-to-zero extrapolation we observe gradual contamination from higher-dimension operators. This contamination seems to be linear and so a flow-time-to-zero-extrapolation is rather straightforward. The resulting zero flow time correlator is then fully renormalized and the leading contribution to the heavy quark momentum diffusion coefficient can be extracted through spectral reconstruction methods. 

For the color-magnetic correlator life is not as easy: it has a nonvanishing anomalous dimension \cite{Laine:2021uzs} which introduces a logarithmic flow time dependence.
Furthermore, it requires additional renormalization to connect it to the $\langle v^2 \rangle$-corrections of heavy quark momentum diffusion.
These problems are out of the scope of this proceedings article and will be addressed in future studies.

\acknowledgments
The authors acknowledge support by the Deutsche Forschungsgemeinschaft (DFG, German Research Foundation) through the CRC-TR 211 'Strong-interaction matter under extreme conditions'– project number 315477589 – TRR 211. The computations in this work were performed on the GPU cluster at Bielefeld University using \texttt{SIMULATeQCD} \cite{mazur2021,Altenkort:2021fqk}. We thank the Bielefeld HPC.NRW team for their support.

\providecommand{\href}[2]{#2}\begingroup\raggedright\endgroup



\end{document}